\documentclass[aps,pre,reprint,amsmath,amssymb,longbibliography]{revtex4-2}
\usepackage{graphicx,booktabs}
\usepackage[T1]{fontenc}
\usepackage{lmodern}
\usepackage[colorlinks=true,linkcolor=blue,citecolor=blue,urlcolor=blue]{hyperref}
\begin{document}
\title{Endogenous wealth cycles from common destruction, production, and delayed relationship repair}
\author{Shuaihua Zhang}
\author{Zining Wang}
\author{Xiaoke Zhang}
\author{Yanling Wu}
\email{ylwu@ysu.edu.cn}
\author{Jun Li}
\email{ljcj007@ysu.edu.cn}
\affiliation{State Key Laboratory of Metastable Materials Science and Technology,
	Hebei Key Laboratory of Microstructural Material Physics,
	School of Science, Yanshan University, Qinhuangdao, 066004, China}
\date{\today}

\begin{abstract}

Wealth accumulates through production and can be destroyed by hostile interactions. We study whether real losses, continuous production, and delayed relationship repair can jointly sustain repeated wealth depletion--recovery cycles in an agent-based economy. Conflicts deduct the same amount from both participants, production adds an equal amount to every agent, and relationships are persistent, with their repair governed by the delayed mean conflict probability. The model retains the actual wealth stock and does not normalize total wealth, so changes in total wealth are determined strictly by the relative strength of realized destruction and production. This exact balance identifies the aggregate turning condition of the cycle, while delayed relationship feedback provides an internal time scale that organizes conflict escalation, wealth depletion, repair, and production-driven rebuilding in sequence. Numerical results indicate that this mechanism produces repeated cycles in the delayed regime studied here. Within a cycle, relative inequality falls during depletion and rises during recovery. Paired production--destruction interventions separate direct loss from subsequent rebuilding, and a finite-population Gini result provides an analytic benchmark for the direct loss response. These results show that a minimal set of rules can sustain wealth cycles at both aggregate and distributional levels, providing a mechanism for endogenous crisis--rebuilding that does not rely on external forcing.

\end{abstract}
\maketitle
\section{Introduction}

Wealth accumulates through production and can be destroyed by hostile interactions. If conflict intensity varies in time, can the competition between destruction and production self-organize into repeated depletion--recovery episodes? We focus not on a single crisis or recovery, but on whether wealth cycles can be sustained as a reproducible, interpretable, and testable collective phenomenon. Asset-exchange models connect individual transaction rules to population wealth distributions \cite{ispolatov1998,dragulescu2000}; saving behavior changes these distributions \cite{chakraborti2000,patriarca2004}, and heterogeneous saving can generate broad tails \cite{chatterjee2004,chatterjee2005}. The yard-sale rule is particularly relevant: the stake is proportional to the wealth of the poorer participant, yet fair winning probabilities can still lead to concentration \cite{caon2007,moukarzel2007}, as explained by kinetic descriptions and Gini monotonicity results \cite{boghosian2014,boghosian2015}; related reviews distinguish these models from descriptions of income and money \cite{chatterjee2007,yakovenko2009}. These works show that individual rules can generate nontrivial distributional dynamics, but conservative or nearly conservative exchange does not produce genuine aggregate depletion. Sustaining wealth cycles therefore requires separating real losses from continuous production.

Allowing wealth to grow and decrease introduces a competition between accumulation and loss. Multiplicative randomness in growth can produce condensation \cite{bouchaud2000}, and nonconservative kinetic models allow uncertain trading returns \cite{cordier2005,during2008}; growth allocation affects concentration and mobility \cite{klein2021,liu2021}, and guaranteed additive income has been studied alongside investment \cite{tobochnik2026}. The harmonic transaction amount has a precedent in conservative exchange \cite{iglesias2012}, and Calvelli and Curado allow both participants to lose that amount together with other trade outcomes, wealth-dependent advantages, taxation, and normalization of total wealth after each Monte Carlo step \cite{calvelli2023}. We retain the actual wealth stock and separate common destruction from equal additive production. This distinction matters for sustaining cycles: without normalization, total wealth genuinely falls when destruction exceeds production and genuinely recovers when production dominates.

Endogenous wealth crises and rebuilding have direct precedents. Benisty obtained inequality-induced crises under mean-wealth normalization \cite{benisty2017}, while Jochim and Bornholdt combined multiplicative growth with dissatisfaction cascades that reset individual wealth, producing collapse, renewed growth, and periodic or quasiperiodic regimes \cite{jochim2025}. Delayed feedback is a known source of collective oscillations \cite{mackey1977,huber2005}. Here the delay acts on relationship repair, while aggregate wealth evolves through production and bilateral destruction. Cycles thus arise from an internal feedback: conflict reinforces relationships, relationships raise future conflict probability, delayed repair reduces conflict, and production rebuilds wealth, without external periodic driving. The delay sequence and population-clock comparison test whether this time scale shares the same origin as the cycles.

The model concerns wealth and relationships only; political institutions and collective action lie outside its rules \cite{epstein2002}. Within this boundary, we show that a simple combination of rules can sustain wealth cycles at both aggregate and distributional levels: an exact wealth balance identifies the aggregate turning condition, delayed relationship feedback provides an internal time scale, paired production--destruction interventions separate direct loss from subsequent rebuilding, and a finite-population Gini result provides an analytic benchmark for the direct loss response and delimits its applicability. Compared with crisis--rebuilding models \cite{jochim2025} and nonconservative exchange models \cite{calvelli2023}, our emphasis is not on more complex transaction rules, but on how real losses, continuous production, and internal delayed feedback close within a single model to sustain wealth cycles.

\section{Model and measurements}

\begin{figure*}[t]
	\centering
	\includegraphics[width=0.96\textwidth]{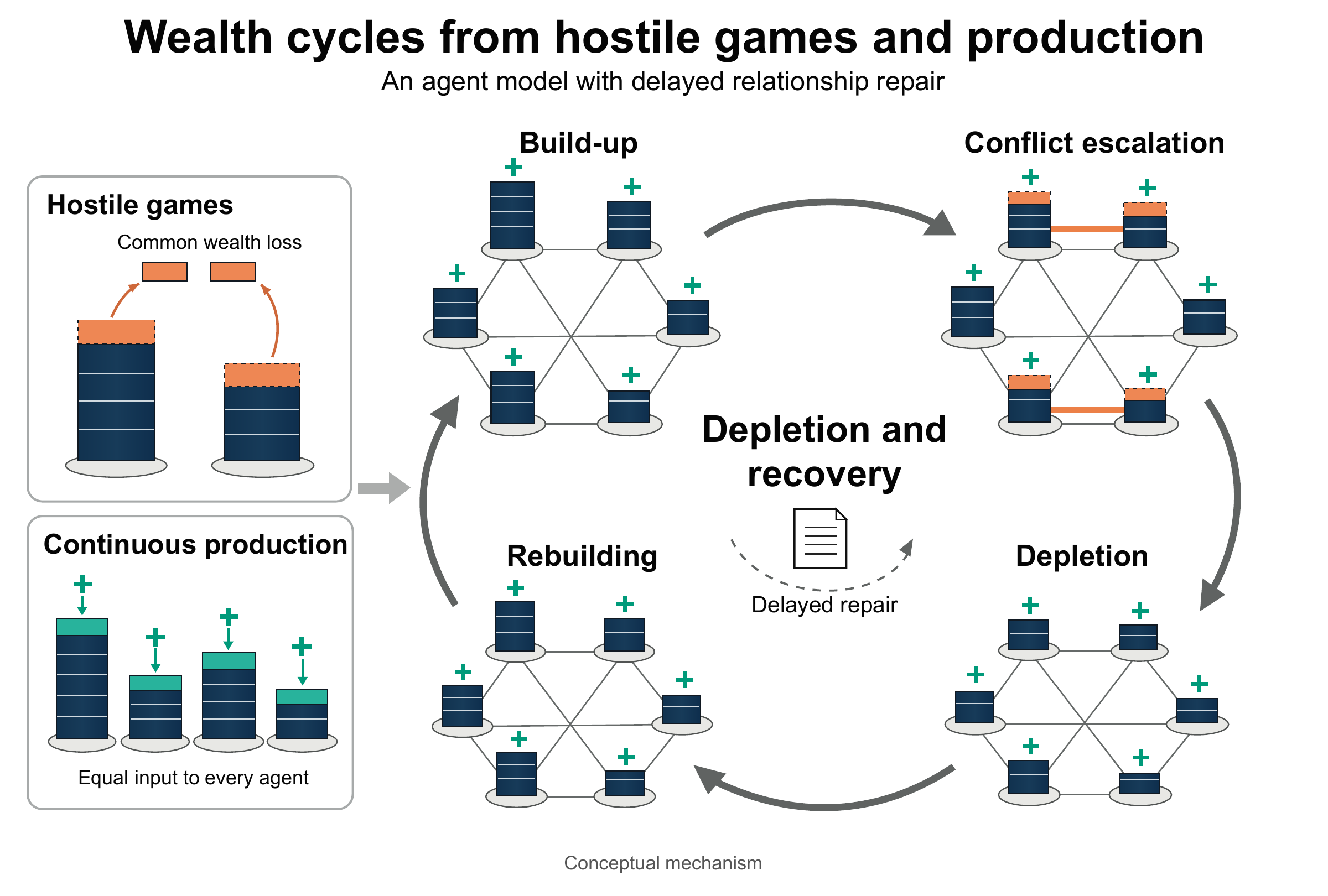}
	\caption{Wealth cycles from hostile games and production. Hostile encounters deduct the same amount from both participants, while equal production adds wealth to every agent. Persistent relationships with delayed repair organize the sequence of wealth accumulation, conflict escalation, depletion, and rebuilding. Stacks and networks are schematic; production operates throughout.}\label{fig:1}
\end{figure*}

Figure 1 summarizes the model and the resulting cycle. There are $N\geq2$ agents with nonnegative wealth $w_i$ and total $W=\sum_iw_i>0$. A round first adds $P$ to every agent and then performs $N$ encounters. Each encounter selects two distinct agents uniformly; repeated visits to the same unordered pair are allowed. A peaceful encounter transfers $r\min(w_i,w_j)$ with equal probability in either direction. A conflict instead deducts the same amount from both participants:

\begin{equation}\label{eq:1}
\begin{aligned}
w_i'&=w_i-\lambda K_\theta(w_i,w_j),\\
w_j'&=w_j-\lambda K_\theta(w_i,w_j).
\end{aligned}
\end{equation}

\begin{equation}\label{eq:2}
\begin{aligned}
K_\theta(x,y)&=(1-\theta)\min(x,y)+\theta\frac{xy}{x+y},\\
&\hspace{35mm}0\leq\theta\leq1 .
\end{aligned}
\end{equation}

The fraction is zero at $x=y=0$. The minimum kernel, $\theta=0$, is the primary model. The harmonic kernel, $\theta=1$, is an analytic comparison: both impose equal bilateral deductions bounded by the poorer wealth, while the smooth rule permits a finite-event inequality theorem. Numerical comparisons test which features survive this change. Both the harmonic amount \cite{iglesias2012} and its common-loss branch \cite{calvelli2023} are inherited rules. Since $K_\theta\leq\min$, fractions $0\leq r,\lambda\leq1$ preserve nonnegative wealth. Numerical fractions are strictly below one, avoiding a degenerate minimum-kernel event that exhausts the total.

Each unordered pair carries a persistent hostility $s_{ij}\in[0,1]$, initially zero. Its conflict probability is $p_{ij}=h+(1-h)s_{ij}$. Let $I_{ij}$ indicate the realized conflict and let $d_{ij}=|w_i-w_j|/(w_i+w_j)$, with zero assigned when both wealths vanish. The visited relationship changes according to

\begin{equation}\label{eq:3}
\begin{aligned}
u_{ij}&=s_{ij}+a(1-s_{ij})I_{ij},\\
s'_{ij}&=(1-b_t)[u_{ij}+\gamma d_{ij}(1-u_{ij})].
\end{aligned}
\end{equation}

All other relationships remain unchanged. The conflict decision and disparity use wealth immediately before that encounter. Thus conflict reinforces the relation, disparity adds hostility, and repair reduces the updated hostility. Repair depends on the past all-pair mean conflict probability,

\begin{equation}\label{eq:4}
\begin{aligned}
q_t&=M^{-1}\sum_{i<j}p_{ij}(t),\qquad M=N(N-1)/2,\\
b_t&=\min[1,b_0+\kappa q_{\max(0,t-\tau)}].
\end{aligned}
\end{equation}

Here $t$ counts completed rounds, $q_0=h$, and repair is fixed during the next round. The quantity $q_t$ averages probabilities over all pairs; it is not the realized fraction of hostile events. Setting $\kappa=0$ makes repair constant, whereas $\gamma=0$ removes current wealth from relationship updates. The feedback is internal, and its explicit memory provides a time scale without periodic external forcing. The remaining parameter ranges are $P\geq0$, $h,a,b_0,\gamma\in[0,1]$, $\kappa\geq0$, and integer $\tau\geq0$.

Transfers conserve total wealth, giving the exact balance

\begin{equation}\label{eq:5}
\begin{aligned}
W_{t+1}-W_t&=NP-D_t,\\
D_t&=2\lambda\sum_{e\in t}I_eK_\theta(w_i^{(e)},w_j^{(e)}).
\end{aligned}
\end{equation}

The loss amount is evaluated at each event's current wealth. We never rescale wealth to impose this identity. Total wealth decreases precisely when realized destruction exceeds production and increases when production exceeds destruction. With $\lambda=0$ and $P>0$, it grows linearly. With $P=0$, it cannot recover. The alternating sign of this balance is the channel through which conflict dynamics produces wealth depletion and recovery.

Relative inequality is measured by the uncorrected finite-population Gini coefficient \cite{cowell2011},

\begin{equation}\label{eq:6}
\begin{gathered}
G=\frac{S}{NW},\qquad S=\sum_{i<j}|w_i-w_j|,\\
G^+=\frac{W}{W+NP}G .
\end{gathered}
\end{equation}

The last identity follows because equal production leaves every absolute difference unchanged. Production therefore reduces Gini more strongly in relative terms when total wealth is depleted. Scaling both initial wealth and production by the same positive factor leaves disparities and relationships unchanged under identical random draws, and scales the entire wealth trajectory.

The delayed comparison uses $N=100$, initial $w_i=100$, $P=1.5$, $h=0.02$, $a=0.15$, $r=\lambda=0.15$, $b_0=0.005$, $\gamma=0.01$, $\kappa=0.416$, and $\tau=1400$. Four shared seeds, 841101--841104, run each $\theta=0,1/2,1$ for 120000 rounds; analysis uses their last halves. A separate constant-repair environment uses $r=\lambda=0.8$, $b_0=0.11$, $\gamma=0.002$, and $\kappa=\tau=0$ for 240000 rounds. These environments differ in more than delay. Additional controls and paired rebuilding experiments are described with their results.

The independent statistical unit is the seed. Blocks, snapshots, pairs, starts, and intervention arms from one seed are dependent observations. Each encounter consumes four pseudorandom draws, including an unused outcome draw on conflict, allowing paired branches to share future draws \cite{glasserman1992}. Wealth and $q$ spectra describe nonoverlapping 20-round means. Spectral periods are reported together with peak recurrence, late-window persistence, and matched controls. Appendix~\ref{app:C} distinguishes the datasets and analysis protocols.

\section{Repeated depletion and recovery}

\begin{figure*}[t]
	\centering
	\includegraphics[width=0.96\textwidth]{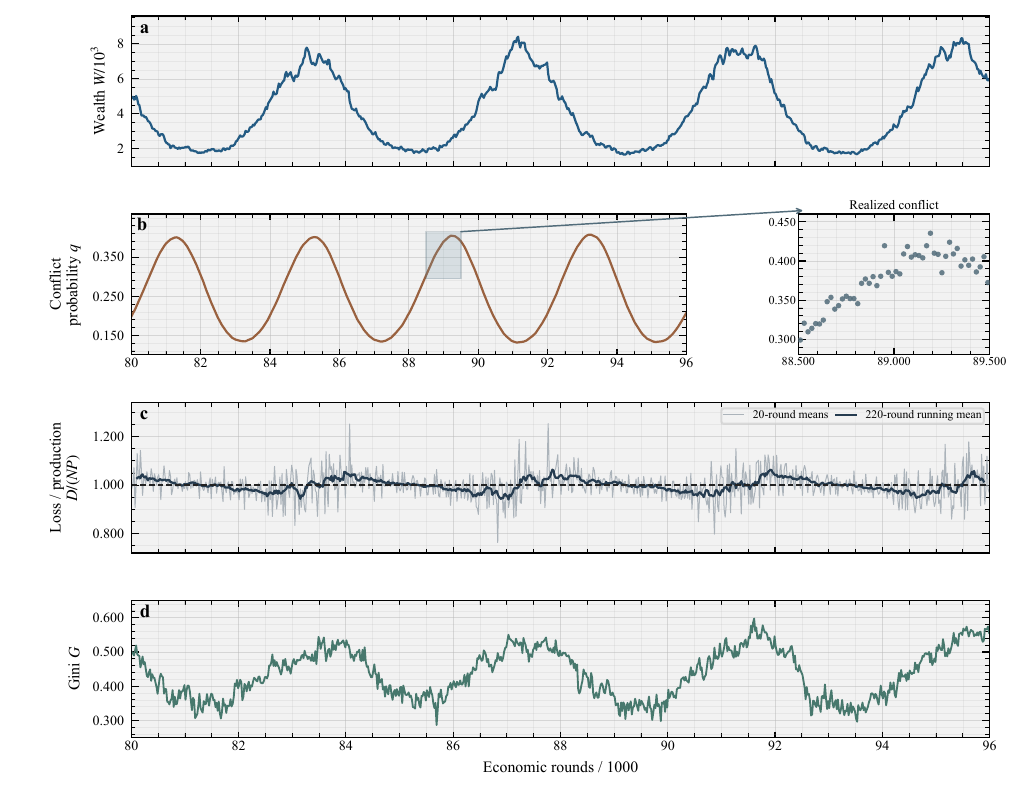}
	\caption{The depletion--recovery mechanism. Minimum-kernel trajectory at the delayed parameters, seed 841101, rounds 80001--96000. The panels show (a) total wealth $W$, (b) all-pair mean conflict probability $q$, (c) destruction relative to production $D/(NP)$, and (d) Gini $G$. Wealth, conflict, and destruction are nonoverlapping 20-round means; Gini is measured at each block end. In (b), the right-hand detail shows the realized conflict fraction in 20-round blocks over rounds 88501--89500. The arrow identifies the corresponding time interval on the main curve, which shows the distinct quantity $q$. The main axis in (b) has its own time scale to accommodate the detail. In (c), the gray trace retains the original block means and the dark curve is a centered 11-block (220-round) running mean used only as a visual guide. The horizontal line at $D/(NP)=1$ separates negative and positive block wealth changes. The block ledger satisfies $\Delta W/20=NP-\overline D$; the wealth curve is a block mean rather than a block endpoint.}\label{fig:2}
\end{figure*}

\subsection{The wealth cycle and its balance}

The delayed-repair environment exhibits recurring depletion and recovery across all three tested kernels. Figure 2 shows the primary minimum-kernel model: increasing conflict drives destruction above production, wealth falls, and the subsequent reduction of destruction permits recovery. The observed balance is exact to numerical precision. Across 56 complete cycles in the four retained minimum-kernel records, total destruction divided by total production is 1.000009; the largest absolute cycle-balance residual is approximately $1.50\times10^{-10}$ wealth units.

For all four seeds and three kernels, wealth and conflict spectra peak near 4000 rounds. Minimum-kernel conflict peaks recur after approximately 3983 rounds (exploratory seed-level 95\% Student-t confidence intervals 3977--3989); late-to-early standard-deviation ratios are 0.988 for conflict and 0.967 for wealth. These measurements support persistent oscillations over the retained 60000-round window.

An earlier independent set of eight seeds, 4001--4008, extends the minimum-kernel model to 210000 rounds each. Its final 105000 rounds contain 25 complete conflict cycles per seed, with mean period 3987.5 rounds. From a low-conflict valley to the next high-conflict peak and then the following valley, mean total wealth follows $8010.5\to1803.5\to8023.1$. Mean Gini follows $0.52784\to0.34699\to0.52837$. Thus depletion accompanies relative equalization, while recovery rebuilds both wealth and concentration. The 200 cycles are nested observations within eight independent runs.

Cycle-aligned mean wealth has a trend of $-0.216\%$ per 10000 rounds (95\% interval $[-1.294\%,0.862\%]$). This measure and the complete-cycle balance characterize recurrence without confusing the phase of a fixed-window endpoint with the cycle mean.

The recurring regime also appears for the harmonic and intermediate kernels. At fixed $\lambda$, mean wealth is 4245.8, 5172.5, and 6608.4 for $\theta=0,1/2,1$. Matching the equal-wealth encounter amount with $\lambda_\theta=0.15/(1-\theta/2)$ gives mean wealth 3651.1 and 2934.4 for $\theta=1/2,1$, respectively.

\subsection{Relationship timing and wealth exposure}

\begin{figure*}[t]
	\centering
	\includegraphics[width=0.96\textwidth]{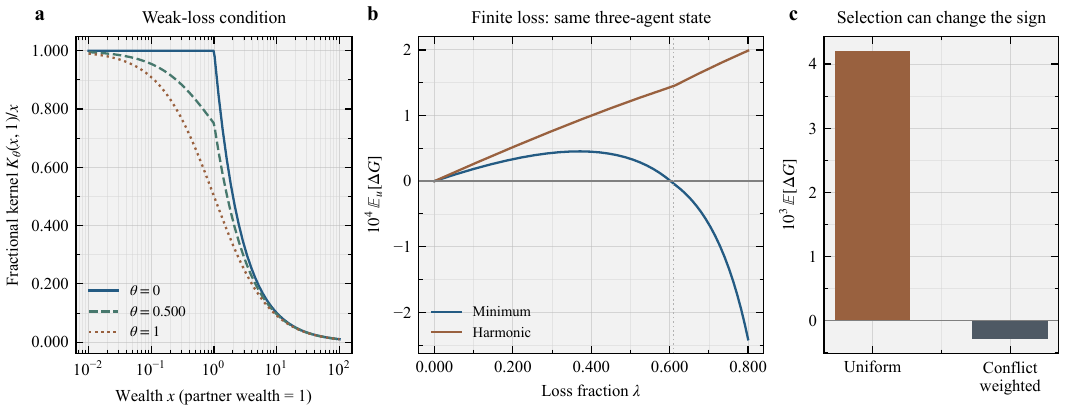}
	\caption{Direct inequality response of common losses. (a) Fractional loss profiles. (b) Uniform minimum and harmonic responses at the constructed state $(0.001,1,1.00061)$. The guide at 0.61 is illustrative, not a sharp threshold. (c) Uniform and conflict-weighted harmonic responses at $(1,10,10)$, $\lambda=0.15$, and probabilities $(0.02,0.02,0.10)$. The uniform theorem does not constrain arbitrary conflict selection.}\label{fig:3}
\end{figure*}
A matched delay sequence resolves the role of relationship memory. For four shared seeds and $\gamma=0.01$, the mean standard deviation of raw $q$ is 0.000989 at $\tau=0$, 0.02966 at 1200, 0.09937 at 1400, and 0.14786 at 1600 (Table~\ref{tab:3}). The respective late-to-early amplitude ratios at 1200, 1400, and 1600 are 0.559, 0.972, and 1.018. The 1200-round signal decays, whereas larger amplitudes persist near and above the main delay over these records. Each exploratory run lasts 60000 rounds; the ratios compare the last and first 10000-round windows of the retained final half. Table~\ref{tab:3} also reports seed dispersion and the $\gamma=0$ sequence.

A relationship changes when visited, at mean rate $\nu=N/M=2/(N-1)$ per round. At fixed pair-time delay $\Theta=\nu\tau\simeq28.28$, populations $N=50,100,200$ give pair-time periods 80.408, 80.480, and 80.628. Oscillations also persist in a constant-disparity diagnostic retaining delayed repair. This diagnostic uses a source calibrated from reference records rather than a matched mean conflict level. These comparisons identify relationship feedback as the relevant timing mechanism, with wealth modifying conflict.

Wealth exposure explains why conflict probability alone does not determine the loss rate. At a frozen state, write $m_{ij}=\min(w_i,w_j)$ and let $\langle\cdot\rangle_u$ denote the uniform unordered-pair average. Since $\min(x,y)=(x+y-|x-y|)/2$,

\begin{equation}\label{eq:7}
\begin{aligned}
\langle m\rangle_u&=\frac{W}{N}\left(1-\frac{NG}{N-1}\right),\\
\mathbb E[D_e\mid\text{state}]
 &=2\lambda\left[q\langle m\rangle_u+\operatorname{Cov}_u(p,m)\right].
\end{aligned}
\end{equation}

Equation~\eqref{eq:7} is an exact single-encounter identity for the minimum kernel. At fixed total wealth, concentration reduces the mean exposed amount; the covariance accounts for relationship-dependent conflict selection. The identity interprets the local loss rate, while Eq.~\eqref{eq:5} connects destruction to aggregate wealth.

The constant-repair strong-loss environment provides a secondary comparison: its large wealth fluctuations have inconsistent peak periods across seeds, while mean conflict varies little. Because several parameters differ from the delayed environment, it characterizes a separate fluctuation regime. The matched delay control above supplies the timing comparison; additional fluctuation diagnostics are summarized in Appendix~\ref{app:D} \cite{lindner2004}.

\section{What distribution is rebuilt?}

\begin{table*}[t]
	\caption{Paired Gini effects at the primary 4000-round horizon. Intervals are individual 95\% Student-t confidence intervals across eight seeds. Destruction effects with and without production are $G_{11}-G_{10}$ and $G_{01}-G_{00}$.}\label{tab:1}
	\centering\small\setlength{\tabcolsep}{3pt}
	\begin{tabular}{@{}lrrlrr@{}}
		\toprule
		Kernel & Start & $\mathcal I_G$ & 95\% interval & With production & Without production \\
		\midrule
		Minimum & 60000 & -0.35883 & [-0.38035, -0.33731] & -0.34460 & +0.01423 \\
		Minimum & 62000 & -0.52990 & [-0.55482, -0.50498] & -0.51653 & +0.01337 \\
		Harmonic & 60000 & -0.36070 & [-0.39015, -0.33125] & -0.34833 & +0.01237 \\
		Harmonic & 62000 & -0.47963 & [-0.51056, -0.44870] & -0.46457 & +0.01505 \\
		\bottomrule
	\end{tabular}
\end{table*}

\begin{figure*}[t]
	\centering
	\includegraphics[width=0.96\textwidth]{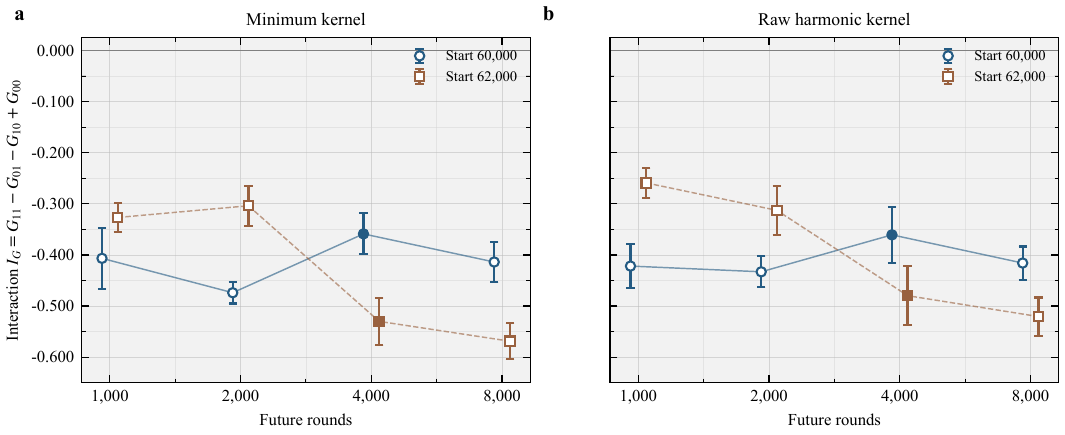}
	\caption{Production--destruction interaction during rebuilding. Mean interaction at each prescribed horizon for minimum and harmonic losses. Colors and markers distinguish fixed starts 60000 and 62000. Filled points mark the primary 4000-round horizon; open points are sensitivity comparisons. Error bars are Bonferroni simultaneous 95\% Student-t confidence intervals over 16 comparisons, each with eight paired seed values. Kernels, starts, branches, and horizons share seeds. The horizontal axis is logarithmic (base 2); lines connect the prescribed horizons as a visual guide.}\label{fig:4}
\end{figure*}

\subsection{Direct common losses and later production}

Destroying the same amount from two agents does not transfer wealth between them. Their mutual difference remains fixed, while distances to other agents and the normalization by total wealth change. Consequently, the direct inequality response must be calculated with the total specific to the selected event. Averaging the denominator first can change the result.

The harmonic comparison makes the direct-loss question analytically tractable. Unlike a trajectory-specific measurement, it gives a statement for every wealth state:

\begin{equation}\label{eq:8}
\begin{gathered}
\mathbb E_u[\Delta G]\geq0,\\
K(x,y)=\frac{xy}{x+y},\qquad 0\leq\lambda\leq1 .
\end{gathered}
\end{equation}

It holds for every $N\geq2$ and every nonnegative wealth profile with positive total. The expectation is over uniform unordered conditional loss pairs. Appendix~\ref{app:A} proves the statement, including ties, rank crossings, and zero wealth. Its physical basis is an ordering of aggregate fractional losses: richer agents lose a smaller fraction on average even after correcting each event for the reduced total. The compensation argument in the appendix makes that ordering valid throughout the legal harmonic loss interval.

Actual conflicts are not uniformly selected when relationships differ. If $F_{ij}$ is the loss-induced Gini increment at the fixed state and $q=\langle p\rangle_u>0$, their conditional mean is

\begin{equation}\label{eq:9}
\begin{aligned}
\mathbb E[\Delta G\mid\mathrm{conflict}]
&=\mathbb E_u F\\
&\quad+\operatorname{Cov}_u(p,F)/q .
\end{aligned}
\end{equation}

For example, harmonic losses at $(1,10,10)$ with $\lambda=0.15$ and probabilities $(0.02,0.02,0.10)$ have positive uniform mean $131/31122$ but negative conflict-weighted mean $-1/3458$. The theorem therefore concerns the loss rule under its stated selection condition, not every relationship-weighted event. For the minimum kernel used in the cycle model and the other tested kernels, we separately enumerate all 4950 pairs at each saved snapshot. Uniform and conflict-weighted means are positive at all 432 snapshots across the nine simulated groups. This numerical check supports the same direct-versus-later distinction in the cycle model; the all-state theorem itself applies to the harmonic rule with uniform conditional selection (Fig.~\ref{fig:3}).

The direct response can coexist with later relative equalization. Equation~\eqref{eq:6} shows why: after destruction lowers total wealth, the same equal production has a larger effect on normalized differences. We test the complete future response rather than assuming that this scale channel is the only mediator.

\subsection{Paired production and destruction cuts}

Eight new seeds, 842101--842108, run the delayed environment with the minimum and harmonic kernels. Complete states at fixed rounds 60000 and 62000 are copied into four branches, retaining wealth, relationships, history, ledgers, and generator state. Binary indices label whether production and destruction are active: $G_{11}$ is the full-model endpoint, $G_{01}$ has no production, $G_{10}$ no destruction, and $G_{00}$ neither. Peaceful trade retains $r=0.15$. A zero-loss conflict still blocks trade and updates the relationship, so removing destruction does not silently replace hostile events with peaceful exchanges.

The interaction is

\begin{equation}\label{eq:10}
\mathcal I_G=G_{11}-G_{01}-G_{10}+G_{00}.
\end{equation}

It compares destruction's later effect with and without production. All future relationships and decisions may respond to the intervened wealth state; this is a comparison of whole rules within the model \cite{holland1986}. Starts and follow-up horizons of 1000, 2000, 4000, and 8000 rounds were fixed before confirmation. The primary horizon is 4000 rounds, giving four primary comparisons, rather than endpoints chosen at each branch's extrema.

All eight seeds have negative interactions in every primary comparison (Table~\ref{tab:1}). On average, destruction raises endpoint Gini without production and lowers it with production. The interaction remains negative at all four prescribed horizons: all 16 kernel/start/horizon comparisons have eight negative seed values and Bonferroni simultaneous 95\% Student-t confidence intervals below zero (Fig.~\ref{fig:4}). These are repeated comparisons from eight independent seeds, not 128 independent experiments. The Student-t confidence intervals assume a suitable sampling distribution of seed-level contrasts. The findings are finite-time results at the specified delayed parameters.

Relative recovery does not imply an absolute benefit from destruction. We follow the lowest 50 identities at each start, with wealth ties broken by identity, distinguishing distributional change from mobility \cite{shorrocks1978mobility,fields1996,cowell2018mobility}. For the minimum kernel at start 60000, their mean cohort total grows from 1040.8 to 3692.8 in the full branch, but reaches 339687.8 without destruction. Thus initially poorer agents can recover from their starting level while remaining much poorer than in the no-destruction comparison. Rank mobility and regression toward the mean also contribute to cohort recovery.

\section{Discussion and conclusion}

Hostile destruction and continuous production produce recurrent wealth depletion and recovery in the studied delayed-repair economy. The wealth balance identifies the turning points: destruction exceeding production lowers the aggregate stock, and production restores it as conflict subsides. The delay sequence and population-clock comparison connect recurrence to the timing of relationship feedback.

The cycles also reorganize relative wealth. Equal production has a stronger normalizing effect when total wealth is low, while subsequent exchanges and conflict selection contribute to the full response. Paired continuations show that destruction lowers later Gini on average with production active and raises it without production, relative to each corresponding no-destruction branch. The finite harmonic-loss theorem and the conflict-selection correction distinguish this later effect from the direct response to a common deduction. Lower Gini describes wealth shares; the initially poorer cohort remains wealthier in absolute terms without destruction.

The model connects individual hostile encounters and production to an aggregate rhythm without a periodic external drive. Its wealth and distribution dynamics complement earlier models of crisis and rebuilding \cite{jochim2025} and nonconservative exchange \cite{calvelli2023}.

\begin{acknowledgments}

This work is funded by Natural Science Foundation of Hebei Province (Grant No. A2026203033), Beijing National Laboratory for Condensed Matter Physics (Grant No. 2024BNLCMPKF020) and Innovation Capability Improvement Project of Hebei Province (Grant No. 22567605H).  The authors used ChatGPT-5.6 to assist with code debugging and language editing. The authors directed its use within the workflow, and all outputs were verified against original data by the authors, who take full responsibility for the content of this work.

\end{acknowledgments}
\begin{table*}[t]
\caption{Provenance of the principal numerical comparisons. The listed seeds are independent within each study. Related branches and cycles remain paired or nested.}\label{tab:2}
\centering\small\setlength{\tabcolsep}{3pt}
\begin{tabular}{@{}p{0.16\textwidth}p{0.16\textwidth}p{0.11\textwidth}p{0.24\textwidth}p{0.25\textwidth}@{}}
\toprule
Study & Seeds & Delay $\tau$ & Rounds per run & Retained observations \\
\midrule
Kernel cycles & 841101--841104 & 1400 & 120000 & Last 60000; 20-round means \\
Long phase records & 4001--4008 & 1400 & 210000 & Last 105000; 25 cycles per seed \\
Delay control & 42--45 & 0 or 1600 & 60000 & Last 30000; raw $q$ \\
Delay sequence & 42--45 & 0--2200; eight values & 60000 & Last 30000; raw $q$; $\gamma=0,0.01$ \\
Paired confirmation & 842101--842108 & 1400 & Start 60000 or 62000, then 8000 & Primary horizon 4000 \\
Population clock & 2001--2004 & 693, 1400, 2814 & 103950, 210000, 422100 & $N=50,100,200$; late records \\
\bottomrule
\end{tabular}
\end{table*}

\appendix
\section{Finite harmonic common-loss inequality}\label{app:A}

Consider a fixed positive wealth state and normalize $x_i=w_i/W$, so $\sum_i x_i=1$. This is an algebraic choice, not a reset of the economy. Sort wealth and let $g_i=2i-N-1$, using midrank values at ties. The vector $g$ is a support gradient of the convex function $S$, and $\sum_i g_ix_i=S(\boldsymbol x)$. For each selected pair define

\begin{equation}\label{eq:A1}
\begin{aligned}
A_{ij}&=\frac{K(x_i,x_j)}{1-2\lambda K(x_i,x_j)},\\
\ell_i^{(\lambda)}&=x_i^{-1}\sum_{j\ne i}A_{ij}.
\end{aligned}
\end{equation}

For the selected pair $(i,j)$, write $x'_k=x_k-\lambda K(x_i,x_j)(\delta_{ki}+\delta_{kj})$. These post-loss amounts retain the pre-event wealth unit: $\sum_k x'_k=1-2\lambda K(x_i,x_j)$. Thus $G'=S(\boldsymbol x')/[N(1-2\lambda K(x_i,x_j))]$. Keeping this event-specific total in Gini gives

\begin{equation}\label{eq:A2}
\begin{aligned}
\mathbb E_u\Delta G
&=-\frac{\lambda}{NM}\operatorname{Cov}_{x}(g,\ell^{(\lambda)})\\
&\quad+\frac{1}{NM}\sum_{i<j}\frac{Q_{ij}}{1-2\lambda K(x_i,x_j)} .
\end{aligned}
\end{equation}

where $Q_{ij}=S(\boldsymbol x')-S(\boldsymbol x)+\lambda K(x_i,x_j)(g_i+g_j)\geq0$. Here the covariance uses weights $x_i$, and the nonnegative remainder includes rank crossings. It remains to show that the corrected fractional rate decreases with wealth.

For the harmonic kernel, $K(x,y)\leq(x+y)/4\leq1/4$, so at least half the total remains. Write $d_{ab}=a+b-2\lambda ab>0$. Then $A_{ij}=x_ix_j/d_{x_ix_j}$ and $\ell_i^{(\lambda)}=\sum_{j\ne i}x_j/d_{x_ix_j}$. For $x=x_i<y=x_j$,

\begin{equation}\label{eq:A3}
\begin{aligned}
\frac{\ell_i^{(\lambda)}-\ell_j^{(\lambda)}}{y-x}
&=\frac{1}{d_{xy}}\\
&\quad+\sum_{k\ne i,j}\frac{z_k(1-2\lambda z_k)}{d_{xz_k}d_{yz_k}} .
\end{aligned}
\end{equation}

All terms are nonnegative for $\lambda\leq1/2$. At larger legal losses, a negative third-agent term requires $z>1/(2\lambda)\geq1/2$. At most one agent can satisfy this. Put $a=1-2\lambda z<0$ and use $x+y\leq1-z$. The positive mutual term compensates that possible negative contribution because

\begin{equation}\label{eq:A4}
\begin{aligned}
B&=d_{xz}d_{yz}-z(2\lambda z-1)d_{xy}\\
 &=z^2+2za(x+y)+xy\,a(1-4\lambda z)\\
 &\geq z^2-2z(2\lambda z-1)(1-z)\\
 &\geq z^2-2z(2z-1)(1-z)\\
 &=z(4z^2-5z+2)>0.                                    
\end{aligned}
\end{equation}

The dropped product is nonnegative because both $a$ and $1-4\lambda z$ are negative. The next inequality uses $\lambda\leq1$, and the final quadratic has discriminant $-7$. Thus the corrected rate is nonincreasing, while rank increases. Their weighted covariance is nonpositive, proving Eq.~\eqref{eq:8}. Midrank support gradients cover ties. The positive post-event bound allows continuity to include zero wealth. Individual pair increments can still be negative; the proof concerns their uniform mean.

\section{Relation to other exchange and distribution mechanisms}\label{app:B}

The harmonic amount appears in conservative exchange \cite{iglesias2012}, and Calvelli and Curado include its common-loss branch within a process that also permits gains and normalizes total wealth \cite{calvelli2023}. Our theorem conditions on one such loss at a fixed finite state. This differs from the fairness assumptions in unbiased binary exchange and its nonconservative extensions \cite{cardoso2021,cardoso2023}. In particular, the latter extension uses a positive conditional return factor common to all agents. For the present uniform harmonic loss rule, at $w_i>0$,

\begin{equation}\label{eq:B1}
\frac{\mathbb E_u[w_i'\mid\boldsymbol w]}{w_i}
=1-\frac{\lambda}{M}\sum_{j\ne i}\frac{w_j}{w_i+w_j}.
\end{equation}

This factor generally depends on $i$. At $(1,2,3)$, the coefficients of $\lambda$ are $17/36$, $14/45$, and $13/60$, so a common return factor is absent for $\lambda>0$. Normalizing after each event also retains a pair-dependent denominator; it cannot be replaced by the mean remaining wealth. The yard-sale Boltzmann Gini theorem already treats finite conservative exchanges \cite{boghosian2015}. Here the distinct calculation is the ordering of the corrected fractional losses in Eqs. (A3)--(A4), including the possible negative third-agent term at large losses. Convexity and fractional-rate ordering are shared mathematical tools, related to progression and Lorenz ordering \cite{jakobsson1973}. Bounds on oligarchy and dynamical Lorenz equations address other distributional questions \cite{cohen2024,cohen2025}.

Probabilistic yard-sale analyses \cite{bouleau2017} and Gamma equilibria of immediate exchange \cite{vanginkel2016} depend on their own event rules. Nonconservative local Nash models provide a further distinct closure \cite{degond2014}. Bounded retention here also differs from affine multiplicative mechanisms with multipliers above one, which can generate power laws \cite{sornette1998}. These comparisons delimit the present mechanism rather than identify its stationary distribution.

Rebuilding through new production is distinct from returning taxes or regulating exchange \cite{cardoso2020,iglesias2023}. Wealth-attained advantage changes transaction bias \cite{boghosian2017}; affine wealth models allow negative wealth \cite{li2019}; growth-allocation and social-protection comparisons change the source or distribution of new wealth \cite{villafane2025}. Network-limited exchanges introduce additional population and topology effects \cite{lee2023}. These operations suggest useful extensions of the present complete-pair model. As with agent-based modeling more generally, interpretation follows from explicit rules \cite{bonabeau2002}.

\section{Numerical protocol}\label{app:C}

\begin{table*}[t]
	\caption{Conflict amplitude and persistence across the delay sequence. All other parameters follow Sec. II. For each run, $\sigma_q$ is the population standard deviation of raw $q$ over rounds 30001--60000, and $R_q$ is the ratio of standard deviations over rounds 50001--60000 and 30001--40000. Entries are means $\pm$ sample standard deviations across four seeds (42--45), paired across parameter cells. The uncertainty describes seed dispersion. No detrending or block averaging is used.}\label{tab:3}
	\centering\small\setlength{\tabcolsep}{3pt}
	\begin{tabular}{@{}rrrrr@{}}
		\toprule
		$\tau$ & $\sigma_q$, $\gamma=0$ & $R_q$, $\gamma=0$ & $\sigma_q$, $\gamma=0.01$ & $R_q$, $\gamma=0.01$ \\
		\midrule
		0 & $0.00107\pm0.00005$ & $0.839\pm0.270$ & $0.00099\pm0.00007$ & $0.803\pm0.245$ \\
		600 & $0.00175\pm0.00021$ & $0.764\pm0.207$ & $0.00186\pm0.00022$ & $0.704\pm0.199$ \\
		900 & $0.00202\pm0.00020$ & $0.818\pm0.519$ & $0.00221\pm0.00053$ & $0.879\pm0.571$ \\
		1200 & $0.00401\pm0.00084$ & $0.612\pm0.192$ & $0.02966\pm0.00224$ & $0.559\pm0.066$ \\
		1400 & $0.03439\pm0.00349$ & $0.828\pm0.057$ & $0.09937\pm0.00190$ & $0.972\pm0.021$ \\
		1600 & $0.07881\pm0.00111$ & $1.004\pm0.030$ & $0.14786\pm0.00042$ & $1.018\pm0.012$ \\
		1800 & $0.10580\pm0.00093$ & $0.951\pm0.015$ & $0.18125\pm0.00071$ & $1.022\pm0.010$ \\
		2200 & $0.14440\pm0.00133$ & $0.939\pm0.008$ & $0.22281\pm0.00076$ & $0.976\pm0.001$ \\
		\bottomrule
	\end{tabular}
\end{table*}

Table~\ref{tab:2} summarizes the independent studies and their retained observations.

The delay sequence contains 64 exploratory runs: eight delays, two disparity responses, and four seeds, each for 60000 rounds. Sixteen runs at $\tau=0,1600$ are the earlier controls; 48 were added in the original exploration. These are separate from the longer confirmation studies. Table~\ref{tab:3} reports all tested cells; $R_q$ measures amplitude persistence, with recurrence assessed separately in Sec. III.

For the reported kernel spectra, we center each complete final-half 20-round block series and apply a symmetric Hann window. Candidate natural Fourier bins have periods of at least 200 rounds and at least three cycles per record. The three-bin fraction sums the selected peak and its adjacent bins, divided by power at all nonzero frequencies. These statistics describe the original kernel study. The later cycle reanalysis uses peak timing and window persistence; its separately computed unwindowed spectral fractions are not pooled with the Hann-window statistics.

The earlier eight-seed phase analysis uses 100-round mean $q$, peak/trough prominence at least 5\% of mean $q$, and minimum spacing 2000 rounds. Between adjacent troughs it selects the highest detected peak; wealth and Gini do not select phases. The four-seed reanalysis uses 20-round means with the same $q$ prominence and spacing. Wealth peaks instead require prominence 10\% of mean wealth. Complete $q$ peak-to-peak intervals define cycle-mean wealth, regressed against cycle midpoints; five equal tail windows define persistence ratios.

The kernel study retains twelve physical snapshots per run: four seeds in each of nine groups, comprising 36 runs. Each snapshot enumerates all 4950 pairs. Source files remain linked to their original run manifests; reanalysis and replay add no independent physical realizations.

\section{A fluctuating constant-repair comparison}\label{app:D}

For the strong-loss constant-repair comparison with $\gamma=0$, write the exact round balance as $W_{t+1}=R_t(W_t+NP)$ with $0\leq R_t\leq1$. An empirical i.i.d. retention background has mean 20-round wealth CV 0.110, compared with 0.545 in four held-out physical runs. The contrast rejects that particular independent-loss amplitude explanation, conditional on the frozen empirical retention pool. It tests amplitude against a specified background, rather than periodicity \cite{theiler1992}. State-dependent exposure and temporal correlations remain part of the physical process. The full pool, bounded-retention derivation, and simulations are retained in the research archive.

\bibliography{references}
\end{document}